%% file: sbgcinf1a.tex
\documentclass[letter,traditabstract]{aa}  
\usepackage{mathptmx}
\usepackage{amssymb}
\usepackage{natbib}
\usepackage{graphicx,color}
\usepackage{txfonts}
\input{def}

\newcommand{\rem}[1]{}
\newcommand{\rev}[1]{{{ #1}}}
\begin{document}
\sloppy
   \title{Superbubble dynamics in globular cluster infancy}
\titlerunning{Superbubbles in globular cluster infancy I}
   \subtitle{I. How do globular clusters first lose their cold gas?}

   \author{Martin Krause \inst{1,2,3}
     \fnmsep\thanks{E-mail: Martin.Krause@universe-cluster.de}
          \and Corinne Charbonnel \inst{3,4}
          \and Thibaut Decressin \inst{3} 
         \and \\ Georges Meynet \inst{3}
          \and Nikos Prantzos  \inst{5}
         \and Roland Diehl \inst{2,1}
        }

   \institute{Excellence Cluster Universe, Technische Universit\"at M\"unchen,
     Boltzmannstrasse 2, 85748 Garching, Germany
         \and
         Max-Planck-Institut f\"ur extraterrestrische Physik, 
         Postfach 1312, Giessenbachstr., 85741 Garching, Germany
         \and
         Geneva Observatory, University of Geneva, 51 Chemin des Maillettes, 1290 Versoix, Switzerland
         \and
IRAP, UMR 5277 CNRS and Universit\'e de Toulouse, 14 Av. E. Belin, 31400 Toulouse, France
\and
         Institut d'Astrophysique de Paris, UMR7095 CNRS, Univ. P. \& M. Curie, 98bis Bd. Arago, 75104 Paris, France
     }

   \date{Received August 2012; accepted }

 
  \abstract
   {The picture of the early evolution of globular clusters
     has been significantly revised in recent years. Current scenarios
   require at least two generations of stars of which the first generation
   (1G), and therefore also the protocluster cloud,
   has been much more massive than the currently predominating second
   generation (2G). Fast gas expulsion is thought to unbind the majority of
 the 1G stars. Gas expulsion is also mandatory to remove
 metal-enriched supernova
ejecta, which are not found in the
2G stars. It has long been thought that the supernovae
themselves are the agent of the gas expulsion, 
based on crude energetics
     arguments. Here, we assume that gas expulsion happens via the formation of a
     superbubble, and describe the kinematics by a thin-shell model.
   We find that supernova-driven shells are destroyed by the
     Rayleigh-Taylor instability before they reach escape speed for
     all but perhaps the least massive and most extended clusters. 
     More power is required to expel the gas, which might plausibly be
     provided by a coherent onset of accretion onto the stellar
     remnants. 
The resulting kpc-sized
   bubbles might be observable in Faraday rotation maps 
   with the planned Square
   Kilometre Array radio telescope against polarised background radio
   lobes if a globular cluster would happen to form in front of such a radio
   lobe.
   }

   \keywords{(Galaxy:) globular clusters: general -- ISM: bubbles --
     ISM: jets and outflows}

   \maketitle
%

\section{Introduction}\label{intro}

Galactic globular clusters (GCs) today typically consist of old 
low-mass stars and little or
no gas. However, there must have been a time when they formed as
gas-rich objects with numerous formation 
of also massive young stars. Many details of this early
epoch have only recently been discovered \citep[][for recent
reviews]{GCB12,Charb10}.
Progress has in particular been made via
spectroscopy and chemical and dynamical evolution modelling. This has
led to a picture of star formation in multiple episodes:
In summary \citep[e.g.][and references therein]{PC06}, the stars in individual 
GCs are mono-metallic regarding the iron group
elements (Fe, Ni, Cu), and have little scatter and similar trends as field
stars for the neutron capture (Ba, La, Eu) and the alpha-elements (Si, Ca).
However, light elements present strong variations from star to star with
 anti-correlations, between O~and~Na, and Mg~and~Al, respectively. 
The interpretation is that GCs form from uniformly
pre-enriched gas, which explains the similarities for the iron group,
neutron capture, and alpha-elements. To explain the
anti-correlations, one requires processed material that has been
subject to hydrogen burning at about 75~MK \citep{Prantzea07}. These conditions
are found in the most massive fast rotating massive stars (FRMS)
and in massive
asymptotic giant branch (AGB) stars. Thus, one requires a first
generation (1G) of stars including massive stars, the ejecta of which form a
second generation (2G) containing low-mass stars (typically the majority of
the stars we observe today). The stellar ejecta have to be
mixed to a varying degree of about 30-50~\% with pristine gas to
produce the abundance patterns (anti-correlation) of the now observed 2G
stars, but the inclusion of processed gas ejected in
supernovae  (SNe) has to be avoided \citep{Prantzea07,Decrea07a,DErcolea11}. 
Gas expulsion by these SNe seemed to be an obvious way to
remove their ejecta from the GC together with the bulk of the gas.
Both the
AGB  and the FRMS scenario agree that the first generation of stars,
and thus the initial total stellar population, was much more massive than the second generation. If the initial mass
function (IMF) would have been normal, many of the 1G
stars would then have had to be lost \citep{Decrea07b,Vespea10,SC11}. 
Assuming mass segregation and
formation of the 2G
stars in the vicinity of the more tightly bound massive 1G stars,
a quick change of the gravitational potential may unbind the major part of the
1G low mass stars in the outskirts of a GC.

Winds and SNe produce interstellar bubbles,
as commonly observed in the interstellar medium
\citep[e.g.][]{Churchea06}.
Given the small separations in GCs, they should soon unite and thus
form a superbubble \citep[e.g.][]{Bagea11,Jaskea11}.
GCs are extremely tightly bound systems with half-mass
radii of typically a few, sometimes only one pc \citep{Harris96}.
They are unlikely to have been less concentrated in the
past \citep{Wilkea03}.
Gravity poses a profound obstacle to
escaping superbubbles. Superbubbles first need to build up a high
pressure to lift up the gas. 
Once the half-mass radius is reached, gravity declines quickly, and
the pressure force strongly dominates, which leads to acceleration of
the shell, and thus triggers the Rayleigh-Taylor (RT)
instability. When RT modes of about the bubble size are able to grow,
the shell fragments and releases its internal pressure. This will
favour its fall back into the central part of the cluster. Here we show that this process
prevents gas expulsion by SN feedback in all but the least
massive GCs.
The power released by accretion onto dark remnants could be
sufficient to expel the gas.
%
%

\section{Superbubble formation and gas expulsion}
Once an amount
of energy comparable to the binding energy is liberated, we expect a
superbubble to form. The
evolution of GC superbubbles has been modelled 
by \citeauthor*{BBT91} (\citeyear{BBT91}, \citeyear{BBT95})\footnote{ 
While the 2G formation
scenario in the supershell proposed in these papers did not stand up
to observational scrutiny (e.g. because SN ejecta are now
thought not to be mixed with the gas that forms the 2G stars), the
hydrodynamics is still valid.}. They showed that
the thin shell approximation (compare below) models the superbubble expansion
faithfully. 

\subsection{The thin-shell model}
We
model the superbubble with the spherically symmetric thin-shell
approximation, where the
change of the shell's momentum is simply given by the applied
forces,
\eql{eq:tse}{ \frac{\partial}{\partial t} ({\cal M} v) = p A - {\cal M}
  g\, .}
Here, ${\cal M}= 4\pi \int_0^r \rho_\mathrm{g}(r^\prime) r^{\prime\,2} \,\mathrm{d}r^\prime$ 
is the mass in the shell, with the gas density $\rho_\mathrm{g}$ and
the shell radius $r$; $v$
is the shell velocity, $p$ the bubble pressure, assumed to dominate
over the ambient pressure, $A=4\pi r^2$ the surface area of the
shell and $g$ the gravitational acceleration.
The bubble pressure is $p=(\gamma-1) (\eta E(t)-{\cal
  M}v^2/2)/V$, with the bubble volume $V=4\pi r^3/3$, the energy
injection law $E(t)$, an efficiency parameter $\eta$, and the ratio of
specific heats, $\gamma=5/3$. 

As input to the model, we need to specify the mass profile and the
energy input. Following other recent work \citep[e.g.][]{BCP08,Decrea10}, 
we use a Plummer model for the spatial distribution of gas and stars.
The gas mass inside a radius $r$ is then given by
\eq{{\cal M}(r) = (1-\epsilon_\mathrm{sf})
  \frac{M_\mathrm{tot}r^3}{(r^2+r_\mathrm{c}^2)^{3/2}}\, ,}
where $r_\mathrm{c}= (2^{2/3}-1)^{1/2} \, r_{1/2}\approx r_{1/2}/1.3$ is the core
radius
and $\epsilon_\mathrm{sf}$ the star formation efficiency. For the gravitational
acceleration, we take into account the stars and half of the gas mass
in the shell. This results in
\eq{g=\frac{1+\epsilon_\mathrm{sf}}{1-\epsilon_\mathrm{sf}}
  \frac{G {\cal M}}{2r^2}\, .}

Our standard case is a protocluster of 
$M_\mathrm{tot} =9 \times 10^6~M_\odot$, a half-mass
radius of $r_{1/2}=3$~pc, a star formation efficiency of 1/3, and  
a Salpeter IMF, which should be applicable for more massive
GCs such as NGC~6752 \citep{Decrea10}.

We have tested three scenarios for the energy injection law $E(t)$:
In our {\em standard} scenario,
we assume that all stars with initial masses between 9 and 120~$M_\odot$
explode as SNe. Following \citet{Decrea10}, we assume the
SNe to contribute $10^{51}$~ergs, each, with an efficiency of
$\eta=0.2$. We take into account
stellar winds and assume that stars above 25~\ms~form
3~\ms~black holes after explosion, which have each a suitable local
supply of gas such that accretion adds energy to the gas at a rate of
20~\% of the Eddington luminosity. The stars between 10 and
25~$M_\odot$ are assumed to form 1.5~$M_\odot$ neutron stars, which
also contribute 20~\% of their Eddington luminosity.
In a second scenario ({\em no BH, late SN}), we assume that stars with
initial masses $>25$~\ms~do not explode, but directly form black holes
\citep[compare][and references therein]{Decrea10}. The only
energy sources are now SNe originating from stars with $M<25M_\odot$.
The third scenario ({\em dark remnant accretion}) assumes sudden accretion onto
all black holes, which have formed as a result of the stellar
evolution of the stars with $M>25M_\odot$, accompanied by an energy
transfer of 20~\% of the Eddington luminosity to the gas,
and no other energy source. Thus this scenario is applicable to the
epoch when all core-collapse SNe have already taken place.
As a subcase, we add accretion onto all
neutron stars with the same efficiency. 
For these assumptions we integrate Eq.~(\ref{eq:tse}) using
a fourth-order Runge-Kutta method with a sufficient time resolution to
reach numerical convergence.

\subsection{Shell kinematics}
\begin{figure*}
   \centering
   \includegraphics[width=.47\textwidth]{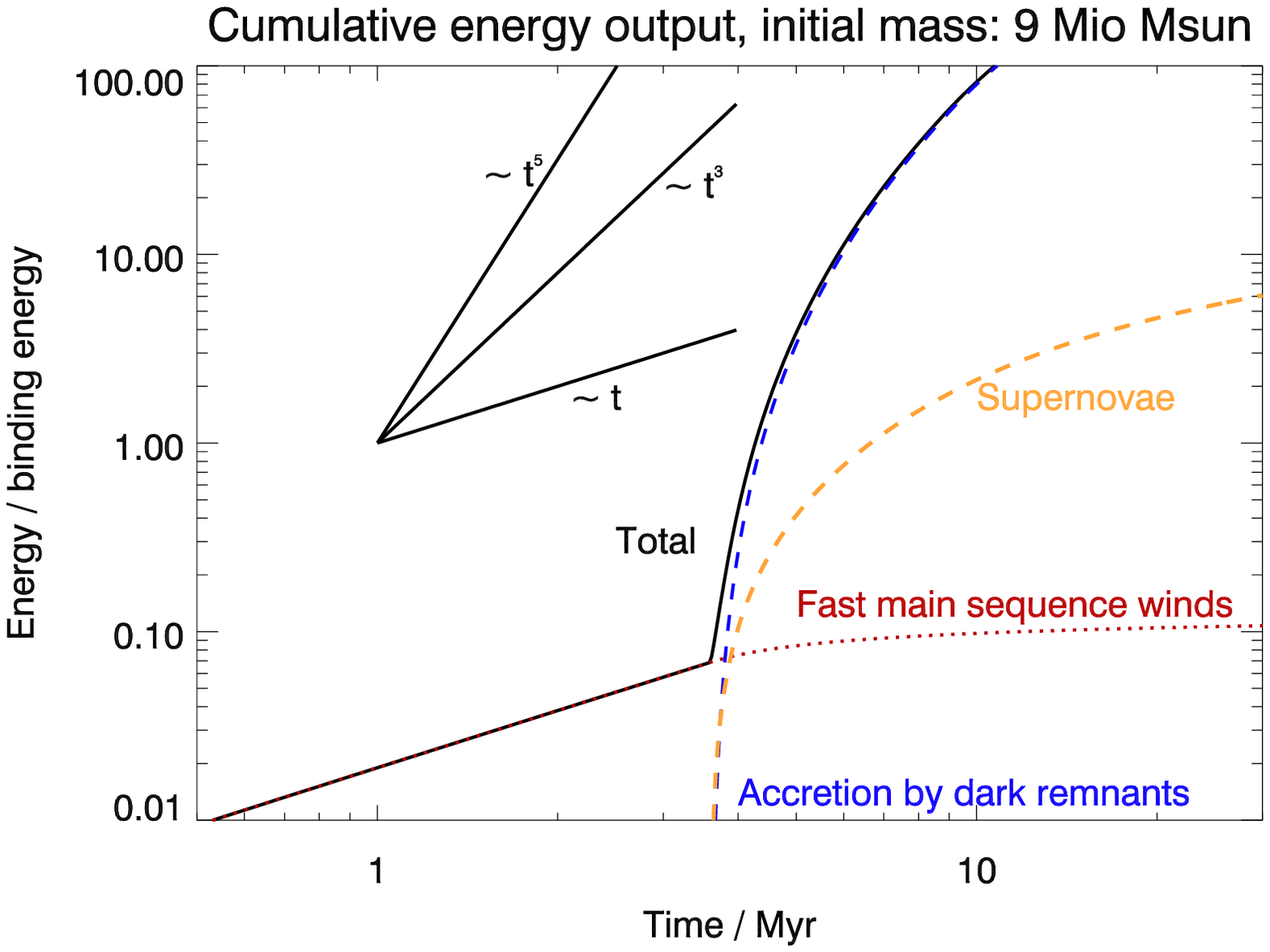}
   \includegraphics[width=.47\textwidth]{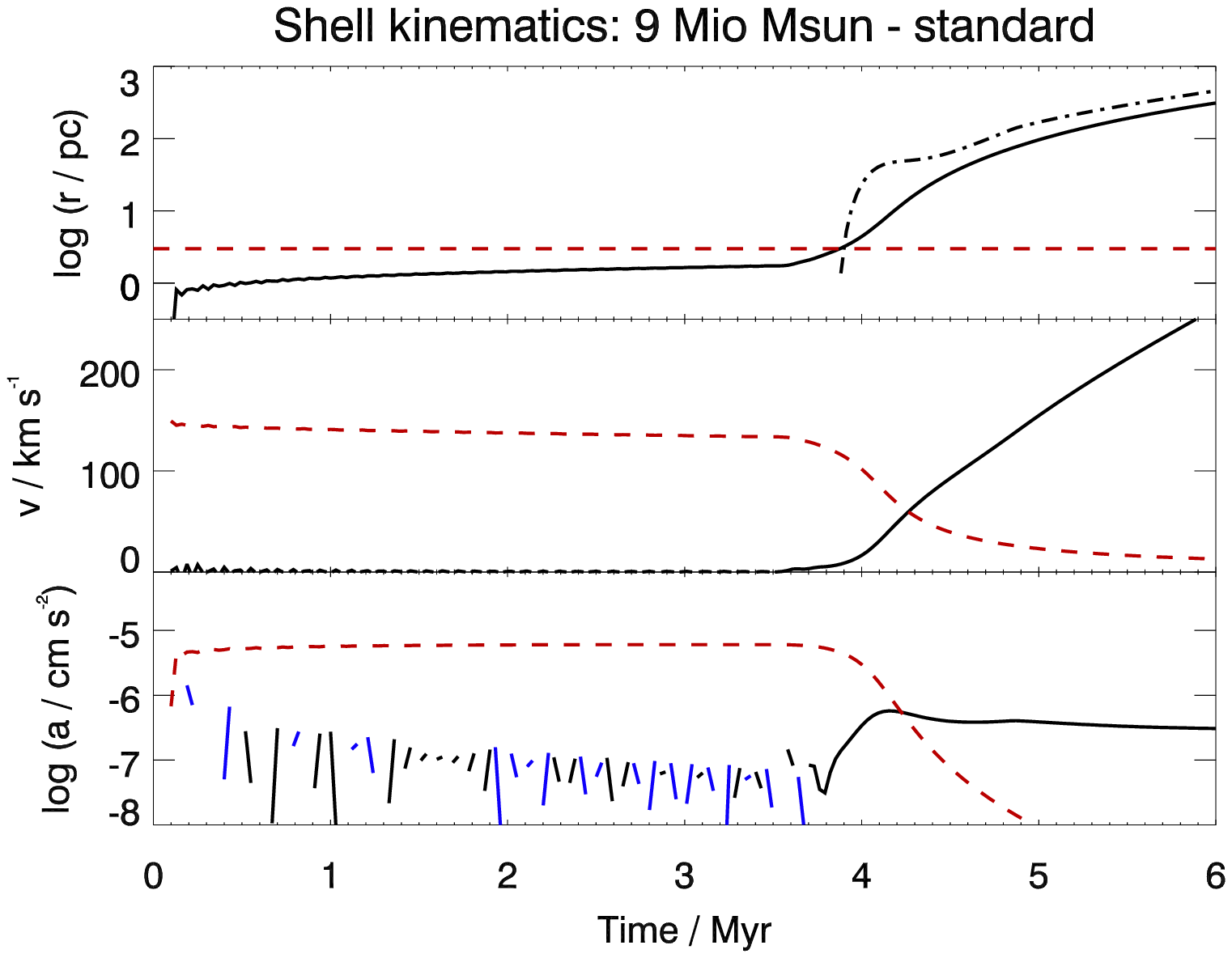}
  \includegraphics[width=.47\textwidth]{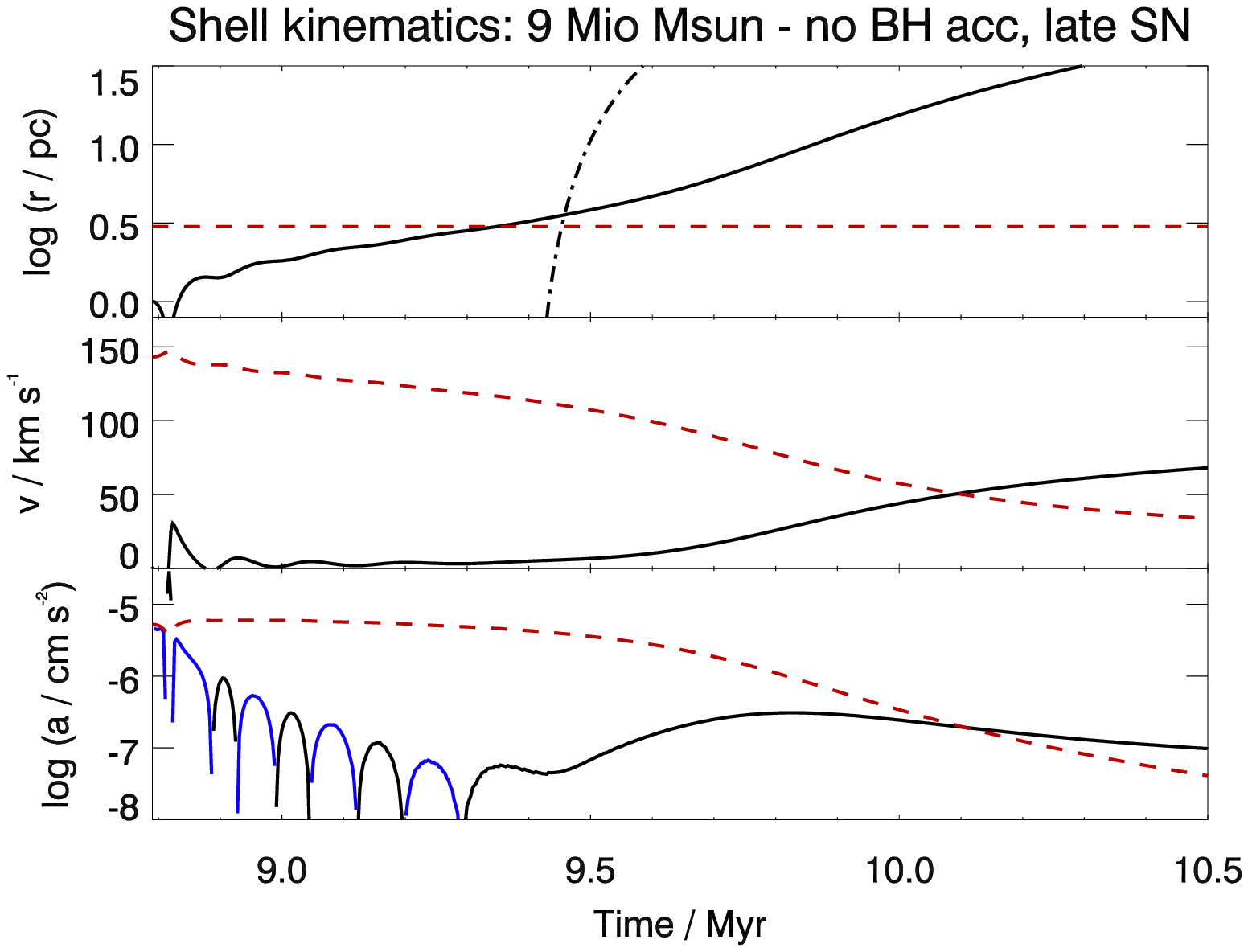}
  \includegraphics[width=.47\textwidth]{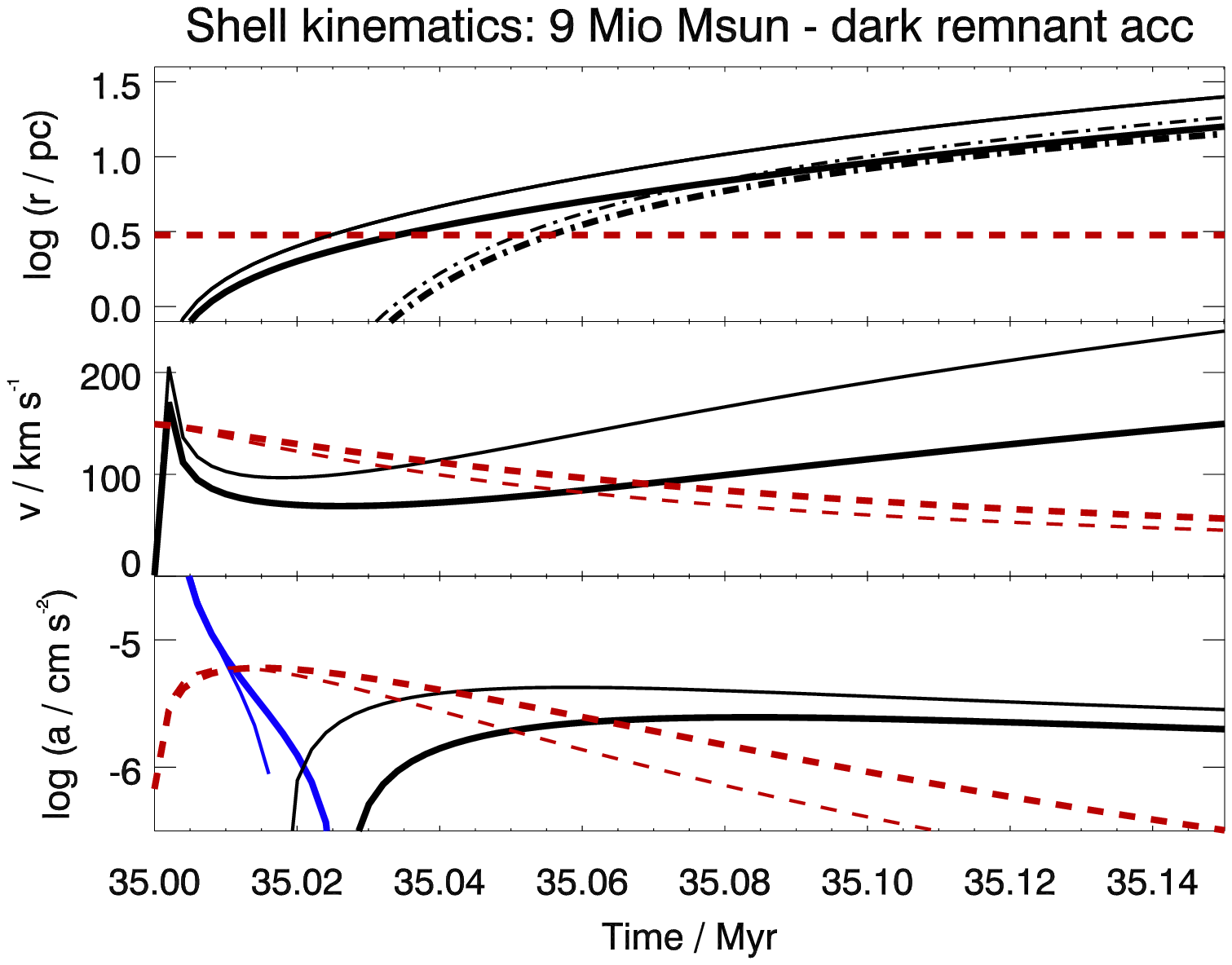}
      \caption{\tiny Produced energy \rev{for the standard scenario} 
        (top left) and superbubble kinematics
        for different assumptions
        about the energy contributors; top right (standard): winds and SNe for
        all massive stars and energy output from all black holes and
        neutron stars; bottom left (no BH acc, late SN): only SNe of
        stars \rev{less massive than 25 $M_\odot$} that 
        explode after \rev{8.79~Myr};
        bottom right (dark remnant acc): only
        sudden accretion onto the dark remnants
        \rev{(thick lines: black holes, only; thin lines: also neutron stars). 
        The timescale for the global evolution of the GC is chosen
        at the birth of a coeval first generation of stars. The abscissae
        indicate the respective starting times of the three considered ejection
        scenarios, which corresponds to the moment when the considered energy
        sources become available. In scenario~3 this could happen any
        time once all the dark remnants have formed,  i.e., after the last SN
        at 35 Myr after the birth of the 1G stars.}
        Within each
        kinematics plot, the upper diagram shows the  bubble radius
        (solid line) and the Rayleigh-Taylor scale (dash-dotted line),
        with the red dashed line indicating the half-mass radius. The
        middle diagram displays the shell velocity (solid line) and
        the escape velocity at the current bubble radius (red dashed
        line). The acceleration (positive: solid black line, negative:
        solid blue line) is shown in the lower diagram, with the
        gravitational acceleration at the current radius shown as a red
        dashed line.}
         \label{fig:sbs}
   \end{figure*}

When neglecting gravity, the
general analytic
solution of the spherically symmetric thin-shell model is known for
arbitrary mass profiles and energy input laws \citep{mypap03a}. 
With power laws for density ($\rho \propto r^{\kappa}$)
and energy injection ($E(t)\propto t^d$), the bubble
expansion law is obtained as 
\eql{eq:bub}{r \propto t^\frac{d+2}{\kappa+5}\, .}
The gravitational pull peaks at $r_\mathrm{c}/\sqrt{2}$, 
and approaches zero for large radii. 
Since the density in the Plummer model drops like $r^{-5}$,
the exponent of $t$ in Eq. (\ref{eq:bub})
approaches infinity in the limit of
large $r$ for any reasonable (Fig.~\ref{fig:sbs}) energy exponent 
$d>-2$.
Hence, we generally expect that the superbubbles are
fairly slow around $r_{1/2}$, where the gravitational pull
is strongest, and quickly accelerate once they have overcome the
gravitational potential well. This triggers the
RT instability. Whenever $a-g>0$, the instability grows
on length scales \citep{Chandra61,BB78}
$\lambda=(a-g)\tau^2$, where $\tau$ is the time the
instability is given to grow. Since $g<0$, only a deceleration
stronger than the gravitational acceleration may stabilise an
expanding shell. This is the case for many standard situations of
interstellar bubbles \citep[e.g.][]{Weavea77}. Yet, acceleration is
unavoidable in the case we consider here.
 In the following, we make the simple assumption that the bubble
has burst and the pressurised hot gas escapes from the cluster, when $\lambda$ reaches
the bubble radius. The shell material
would then collapse back into the cluster, unless it has already
reached escape speed.  We take the time during which the
bubble shows significant acceleration to define $\tau$.

We show the energy injection together with the resulting bubble
kinematics for the three respective assumptions about the energy
injection law in Fig.~\ref{fig:sbs}. 
In the {\em standard case} (all
energy sources active),
the shell expands very slowly to about 4~Myr and in an oscillatory 
manner, as seen from
the alternating sign of the acceleration. Once it reaches
$r_{1/2}$, the acceleration increases strongly and stays at a high
level throughout, 
as expected from the above analysis. The shell reaches the local
escape speed (red dashed line in the middle panels) at about
4.3~Myr, which is too late to avoid pressure loss due to the
RT instability. The cluster's gas, which is now in the
shell fragments, should therefore remain bound to the cluster.
For the second scenario ({\em no BH, late SN}), the shell starts to
fast accelerate after about 0.5~Myr after the SN activity
is assumed to start. However, the escape speed is
reached only after 1.2~Myr. The shell is RT unstable
long before, and therefore this scenario does not lead to gas
expulsion, either. 
The situation is different for the third scenario, {\em dark remnant accretion}:
Here, the shell reaches escape speed immediately, due to the sudden
power increase. The velocity drops slightly because of the hydrodynamic
evolution of the bubble. The escape speed is reached again after only
0.06~Myr and even 0.03~Myr if one includes the neutron stars. The
RT instability is not able to affect the entire shell,
and consequently, the gas is expelled from the cluster.
We have also investigated cases with initial masses 
$M_\mathrm{tot}=10^6$~\ms~and 
$M_\mathrm{tot}=2\times 10^7$~\ms~with very similar results.
The only difference is that for the
high-mass cluster, gas expulsion by dark remnant accretion 
only works with the help of the neutron stars.
The crossing time for the model clusters $\tau_\mathrm{c} = 2.82 r_{1/2}^{3/2}
(GM_\mathrm{tot})^{1/2}$, is 0.22, 0.05, and
0.034~Myr, for the 1, 9, and 20 $\times10^6$~\ms~
clusters,
respectively, using the definition of
\citet{Decrea10}. For all our dark remnant accretion cases with the
exception of the black-hole-only  
case for the high-mass cluster, the shells
reach the half-mass radius much faster,
and we may expect that the outer 1G stars will
also be lost \citep[compare][]{Decrea10}.
We have investigated the parameter space in
$\epsilon_\mathrm{sf}$ and $r_{1/2}$. For both parameters, we find
critical values, above which gas expulsion by SNe only, or by
the power sources in our standard scenario would be possible. 
They are excessive, apart from perhaps the lowest mass
case, where a GC may lose its gas at
$\epsilon_\mathrm{sf} >47$~\%. Yet, this value is too high to expell a
significant number of 1G stars \citep[][]{Decrea10}. Similarly at
$\epsilon_\mathrm{sf}=0.33$,  the SNe succeed only for
$r_{1/2}>4$~pc, which is on the high end of the observed
values.

%
\begin{table}
\caption{Minimum star formation efficiency and half-mass radius for
  SN feedback to be able to expel the gas}             
\label{table:critvals}      
\centering                          
\begin{tabular}{c c c c}        
\hline\hline                 
$M_\mathrm{tot}$\tablefootmark{a} $/ 10^6 M_\odot$ 
  & $\epsilon_\mathrm{sf,crit,std}$\tablefootmark{b}
  & $\epsilon_\mathrm{sf,crit,sn}$\tablefootmark{c}
  & $r_{1/2,crit}$\tablefootmark{d} / pc \\    
\hline                        
     1 & 0.97 & 0.47  & 4\\      
     9 & $>0.99$ & $>0.99$ &18 \\
   20 & $>0.99$ & $>0.99$ & 35 \\
\hline                                   
\end{tabular}
\tablefoot{ 
\tablefoottext{a}{Total initial gas mass out of which the cluster forms.}
\tablefoottext{b}{Critical star formation efficiency 
for our standard energy injection scenario and a half-mass radius of 3~pc.}
\tablefoottext{c}{Critical star formation efficiency 
for the 'SN only' energy injection scenario
and a half-mass radius of 3~pc.}
\tablefoottext{d}{Critical half-mass radius 
in the 'SN only' scenario, assuming a star
  formation efficiency of 0.33.}}
\end{table}

\section{Gas expulsion powered by dark-remnant accretion 
  and possible observational tests}
Up to now, the gas expulsion scenario via SNe has been central to
the two main scenarios for self-enrichment in GCs
\rev{\citep[e.g.][]{BCP08,DErcolea08,Decrea10}}. Here we show that this does
not generally work for simple and
standard assumptions about gas expulsion via a superbubble. 
While the energy injected by SNe in total is
sufficient, it is not delivered fast enough to overcome the
RT instability.
The result should not be restricted to
the Plummer model: Any GC formation
scenario should involve a strongly concentrated gas cloud. The
gravitational pull will always be strongest on scales
comparable to the half-mass radius. Sudden acceleration,
when the gravitational well is
overcome, and RT instability, are the natural consequences. 
The asymptotic acceleration is particularly strong for the Plummer
model. It is likely to occur, however, albeit at a
weaker level for all reasonable profiles
(compare Eq.~\ref{eq:bub} and Fig.~\ref{fig:sbs}).

The only way to overcome the shell destruction by the RT instability
is to inject the energy sufficiently fast, such that the gravitational
well becomes less important for the dynamics. We show that a sudden
activation of all dark remnants \rev{-- a hypothesis, details of which
need to be worked out in the future -- } is plausibly sufficient for this
purpose. Here, we have adopted a general efficiency factor of
20~\%, mainly for consistency with previous work
\citep{Decrea10}. How may we motivate such an efficiency factor for
dark remnant accretion? \rev{The Bondi accretion rate, $10^{-4} M_3^2
  n_6/v_1^3 M_\odot/\mathrm{yr}$, 
$M_3$ being the remnant mass in units of $3 M_\odot$,
  $n_6$ the ISM number density in $10^6$~cm$^{-3}$ and $v_1$ the higher
  of relative velocity and ISM sound speed in km/s, may exceed the Eddington
accretion rate by a large factor whenever the star is close to its outer turning
point. Assembling mass to its vicinity in this phase, a remnant may
plausibly be activated for a large part of the orbit.}
Accretion onto compact stellar sized objects usually
leads to emission in the X-ray part of the spectrum
\citep[e.g.][]{Chiangea10}. For our standard model cluster, 
the hydrogen column density is about
$N_\mathrm{H} \approx 10^{25}$~cm$^{-2}$, i.e. it is
Compton-thick and a large part of the radiated energy might be
absorbed. 
\rev{The shells should have densities of about $10^{6-7}$~cm$^{-2}$ and
  accordingly cool down \citep{SD93}, collapse, remain neutral, and 
therefore also absorb  X-rays efficiently.}
Additionally, there might be 
jets, which in the case of supermassive
black holes are also known to come sometimes close to the Eddington
luminosity \citep[e.g.][]{Krause2005a,CG08}. Since the compact objects
in this scenario would not accrete from a binary companion, but from
the general ISM, one may speculate that the jet powers might also be
comparable to the supermassive black hole case. Jets 
communicate their energy efficiently to
their surroundings via radio lobes
\citep[e.g.][]{Gaiblea09}.
\rev{For our dark-remnant accretion scenario, we find a limiting
  initial cloud mass of about $10^7 M_\odot$ above which the
  cold gas may not be ejected and therefore might form additional stars. 
   This might contribute to explanations of the observed differences at the
   high-mass end of GCs \citep[e.g. greater Fe spreads, ][]{Carea10b}.}


It would be challenging to detect the 
dark remnants during their active
phases, as the emission peaks in the
X-ray part of the spectrum and the clusters should be Compton-thick at
that time. The total X-ray luminosity of the cluster should be low, about
$10^{41}$~erg/s, the active time from our calculations is only about
$10^4-10^5$~years, and the prime objects of interest are high-redshift
galaxies. 
If the radio luminosity were similar, 
one would expect fluxes of about $\mu$Jy, well in the
reach of the upcoming Square Kilometre Array radio telescope (SKA)
\citep[e.g.][and references therein]{Krausea09}. One could at least
constrain well-defined models of cluster formation. If
GC-formation was triggered by galactic scale shock waves \citep{HH11}, e.g.
associated with galactic winds and jets
\citep{mypap02a,Krause2005b}, where the GCs are supposed to form in a
galactic wind shell, 
one might expect some active GCs during
the time when the jet is active too, thus marking the relevant
evolutionary epoch of the GCs.
The dark-remnant driven bubbles could easily reach sizes of kpc or
even 10~kpc on timescales of about $10^7$ years. They might then leave
a signature in the Faraday rotation signal if seen against a polarised
background source.
The SKA should also be able to detect high-redshift radio lobes in
polarisation \citep{Krausea09}. The plasma closest to the radio
sources is usually responsible for a big part of the rotation
measure. Changing the structure of this material can leave observable
features in Faraday rotation maps \citep{HEKA11b}. 

\begin{acknowledgements}
\rev{We thank the referee, R. McCray, for comments that helped to improve
this paper.}
 This research was supported by the cluster of excellence ``Origin and
Structure of the Universe'' (www.universe-cluster.de) and  the ESF
EUROCORES Programme 'Origin of the Elements and Nuclear History of the
Universe' (grants 189 and 190).
CC and TD also acknowledge financial support from the Swiss 
National Science Foundation (FNS) and the French Programme 
National de Physique Stellaire (PNPS) of CNRS/INSU.
\end{acknowledgements}

\bibliographystyle{aa}
\bibliography{/Users/mkrause/texinput/references}

\end{document}

%% file: def.tex
\bibpunct{(}{)}{;}{a}{}{,}
\newcommand{\ignore}[1]{}
\newcommand{\eq}[1]{\begin{equation} #1 \end{equation}}
\newcommand{\eql}[2]{\begin{equation} \label{#1} #2 \end{equation}}

\newcommand{\ms}{$M_\odot$}

%% file: sbgcinf1a.bbl
\begin{thebibliography}{34}
\expandafter\ifx\csname natexlab\endcsname\relax\def\natexlab#1{#1}\fi

\bibitem[{{Bagetakos} {et~al.}(2011){Bagetakos}, {Brinks}, {Walter}, {de Blok},
  {Usero}, {Leroy}, {Rich}, \& {Kennicutt}}]{Bagea11}
{Bagetakos}, I., {Brinks}, E., {Walter}, F., {et~al.} 2011, \aj, 141, 23

\bibitem[{{Baumgardt} {et~al.}(2008){Baumgardt}, {Kroupa}, \&
  {Parmentier}}]{BCP08}
{Baumgardt}, H., {Kroupa}, P., \& {Parmentier}, G. 2008, \mnras, 384, 1231

\bibitem[{{Bernstein} \& {Book}(1978)}]{BB78}
{Bernstein}, I.~B. \& {Book}, D.~L. 1978, \apj, 225, 633

\bibitem[{{Brown} {et~al.}(1991){Brown}, {Burkert}, \& {Truran}}]{BBT91}
{Brown}, J.~H., {Burkert}, A., \& {Truran}, J.~W. 1991, \apj, 376, 115

\bibitem[{{Brown} {et~al.}(1995){Brown}, {Burkert}, \& {Truran}}]{BBT95}
{Brown}, J.~H., {Burkert}, A., \& {Truran}, J.~W. 1995, \apj, 440, 666

\bibitem[{{Carretta} {et~al.}(2010){Carretta}, {Bragaglia}, {Gratton},
  {Lucatello}, {Bellazzini}, {Catanzaro}, {Leone}, {Momany}, {Piotto}, \&
  {D'Orazi}}]{Carea10b}
{Carretta}, E., {Bragaglia}, A., {Gratton}, R.~G., {et~al.} 2010, \apjl, 714,
  L7

\bibitem[{{Celotti} \& {Ghisellini}(2008)}]{CG08}
{Celotti}, A. \& {Ghisellini}, G. 2008, \mnras, 385, 283

\bibitem[{{Chandrasekhar}(1961)}]{Chandra61}
{Chandrasekhar}, S. 1961, {Hydrodynamic and hydromagnetic stability}, ed.
  {Chandrasekhar, S.}

\bibitem[{{Charbonnel}(2010)}]{Charb10}
{Charbonnel}, C. 2010, in IAU Symposium, Vol. 266, IAU Symposium, ed. R.~{de
  Grijs} \& J.~R.~D. {L{\'e}pine}, 131--142

\bibitem[{{Chiang} {et~al.}(2010){Chiang}, {Done}, {Still}, \&
  {Godet}}]{Chiangea10}
{Chiang}, C.~Y., {Done}, C., {Still}, M., \& {Godet}, O. 2010, \mnras, 403,
  1102

\bibitem[{{Churchwell} {et~al.}(2006){Churchwell}, {Povich}, {Allen}, {Taylor},
  {Meade}, {Babler}, {Indebetouw}, {Watson}, {Whitney}, {Wolfire}, {Bania},
  {Benjamin}, {Clemens}, {Cohen}, {Cyganowski}, {Jackson}, {Kobulnicky},
  {Mathis}, {Mercer}, {Stolovy}, {Uzpen}, {Watson}, \& {Wolff}}]{Churchea06}
{Churchwell}, E., {Povich}, M.~S., {Allen}, D., {et~al.} 2006, \apj, 649, 759

\bibitem[{{Decressin} {et~al.}(2010){Decressin}, {Baumgardt}, {Charbonnel}, \&
  {Kroupa}}]{Decrea10}
{Decressin}, T., {Baumgardt}, H., {Charbonnel}, C., \& {Kroupa}, P. 2010, \aap,
  516, A73

\bibitem[{{Decressin} {et~al.}(2007{\natexlab{a}}){Decressin}, {Charbonnel}, \&
  {Meynet}}]{Decrea07b}
{Decressin}, T., {Charbonnel}, C., \& {Meynet}, G. 2007{\natexlab{a}}, \aap,
  475, 859

\bibitem[{{Decressin} {et~al.}(2007{\natexlab{b}}){Decressin}, {Meynet},
  {Charbonnel}, {Prantzos}, \& {Ekstr{\"o}m}}]{Decrea07a}
{Decressin}, T., {Meynet}, G., {Charbonnel}, C., {Prantzos}, N., \&
  {Ekstr{\"o}m}, S. 2007{\natexlab{b}}, \aap, 464, 1029

\bibitem[{{D'Ercole} {et~al.}(2011){D'Ercole}, {D'Antona}, \&
  {Vesperini}}]{DErcolea11}
{D'Ercole}, A., {D'Antona}, F., \& {Vesperini}, E. 2011, \mnras, 415, 1304

\bibitem[{{D'Ercole} {et~al.}(2008){D'Ercole}, {Vesperini}, {D'Antona},
  {McMillan}, \& {Recchi}}]{DErcolea08}
{D'Ercole}, A., {Vesperini}, E., {D'Antona}, F., {McMillan}, S.~L.~W., \&
  {Recchi}, S. 2008, \mnras, 391, 825

\bibitem[{{Gaibler} {et~al.}(2009){Gaibler}, {Krause}, \&
  {Camenzind}}]{Gaiblea09}
{Gaibler}, V., {Krause}, M., \& {Camenzind}, M. 2009, \mnras, 400, 1785

\bibitem[{{Gratton} {et~al.}(2012){Gratton}, {Carretta}, \&
  {Bragaglia}}]{GCB12}
{Gratton}, R.~G., {Carretta}, E., \& {Bragaglia}, A. 2012, \aapr, 20, 50

\bibitem[{{Harris} \& {Harris}(2011)}]{HH11}
{Harris}, G.~L.~H. \& {Harris}, W.~E. 2011, \mnras, 410, 2347

\bibitem[{{Harris}(1996)}]{Harris96}
{Harris}, W.~E. 1996, \aj, 112, 1487

\bibitem[{{Huarte-Espinosa} {et~al.}(2011){Huarte-Espinosa}, {Krause}, \&
  {Alexander}}]{HEKA11b}
{Huarte-Espinosa}, M., {Krause}, M., \& {Alexander}, P. 2011, \mnras, 418, 1621

\bibitem[{{Jaskot} {et~al.}(2011){Jaskot}, {Strickland}, {Oey}, {Chu}, \&
  {Garc{\'{\i}}a-Segura}}]{Jaskea11}
{Jaskot}, A.~E., {Strickland}, D.~K., {Oey}, M.~S., {Chu}, Y.-H., \&
  {Garc{\'{\i}}a-Segura}, G. 2011, \apj, 729, 28

\bibitem[{{Krause}(2002)}]{mypap02a}
{Krause}, M. 2002, \aap, 386, L1

\bibitem[{{Krause}(2003)}]{mypap03a}
{Krause}, M. 2003, \aap, 398, 113

\bibitem[{{Krause}(2005{\natexlab{a}})}]{Krause2005b}
{Krause}, M. 2005{\natexlab{a}}, \aap, 436, 845

\bibitem[{{Krause}(2005{\natexlab{b}})}]{Krause2005a}
{Krause}, M. 2005{\natexlab{b}}, \aap, 431, 45

\bibitem[{{Krause} {et~al.}(2009){Krause}, {Alexander}, {Bolton},
  {Geisb{\"u}sch}, {Green}, \& {Riley}}]{Krausea09}
{Krause}, M., {Alexander}, P., {Bolton}, R., {et~al.} 2009, \mnras, 400, 646

\bibitem[{{Prantzos} \& {Charbonnel}(2006)}]{PC06}
{Prantzos}, N. \& {Charbonnel}, C. 2006, \aap, 458, 135

\bibitem[{{Prantzos} {et~al.}(2007){Prantzos}, {Charbonnel}, \&
  {Iliadis}}]{Prantzea07}
{Prantzos}, N., {Charbonnel}, C., \& {Iliadis}, C. 2007, \aap, 470, 179

\bibitem[{{Schaerer} \& {Charbonnel}(2011)}]{SC11}
{Schaerer}, D. \& {Charbonnel}, C. 2011, \mnras, 413, 2297

\bibitem[{{Sutherland} \& {Dopita}(1993)}]{SD93}
{Sutherland}, R.~S. \& {Dopita}, M.~A. 1993, \apjs, 88, 253

\bibitem[{{Vesperini} {et~al.}(2010){Vesperini}, {McMillan}, {D'Antona}, \&
  {D'Ercole}}]{Vespea10}
{Vesperini}, E., {McMillan}, S.~L.~W., {D'Antona}, F., \& {D'Ercole}, A. 2010,
  \apjl, 718, L112

\bibitem[{{Weaver} {et~al.}(1977){Weaver}, {McCray}, {Castor}, {Shapiro}, \&
  {Moore}}]{Weavea77}
{Weaver}, R., {McCray}, R., {Castor}, J., {Shapiro}, P., \& {Moore}, R. 1977,
  \apj, 218, 377

\bibitem[{{Wilkinson} {et~al.}(2003){Wilkinson}, {Hurley}, {Mackey}, {Gilmore},
  \& {Tout}}]{Wilkea03}
{Wilkinson}, M.~I., {Hurley}, J.~R., {Mackey}, A.~D., {Gilmore}, G.~F., \&
  {Tout}, C.~A. 2003, \mnras, 343, 1025

\end{thebibliography}
